# Autoignition and detonation development from a hot spot inside a closed chamber: effects of end wall reflection

Peng Dai [a], Zheng Chen [b], Xiaohua Gan [a], Mikhail A. Liberman [c]

[a] Department of Mechanics and Aerospace Engineering, Southern University of Science and Technology, Shenzhen 518055, China

[b] SKLTCS, CAPT, BIC-ESAT, College of Engineering, Peking University, Beijing 100871, China

[c] Nordita, KTH Royal Institute of Technology and Stockholm University, Roslagstullsbacken 23, SE-10691 Stockholm, Sweden

**Corresponding author:** Peng Dai, Ph.D.
Department of Mechanics and Aerospace Engineering,
Southern University of Science and Technology,
Shenzhen 518055, China
Tel: +86-(755) 8801-0972
Email: daip@sustech.edu.cn

**Paper length (method 1):**

| | | | |
|---|---|---|---|
| Main Text: | word processor count | = | 3389 |
| Equations: | (13 line + 11) × (7.6 words/line) × (1 column) | = | 182 |
| References: | (30+2)×(2.3 lines/reference) × (7.6 words/line) | = | 559 |
| Figure Captions: | word processor count | = | 286 |
| Figure 1: | (42 mm+10) × (2.2 words/mm) × (1 column) | = | 114 |
| Figure 2: | (53.7 mm+10) × (2.2 words/mm) × (3 column) | = | 420 |
| Figure 3: | (53.7 mm+10) × (2.2 words/mm) × (1 column) | = | 140 |
| Figure 4: | (53.7 mm+10) × (2.2 words/mm) × (2 column) | = | 280 |
| Figure 5: | (53.7 mm+10) × (2.2 words/mm) × (1 column) | = | 140 |
| Figure 6: | (53.7 mm+10) × (2.2 words/mm) × (2 column) | = | 280 |
| | **Total** | **=** | **5792 words** |

**Supplementary Material:** Yes (one file on chemical mechanism validation, 0D homogeneous ignition, and typical autoignition cases)

**Color reproduction:** no (all color figures are to be printed in gray scale)

# Autoignition and detonation development from a hot spot inside a closed chamber: effects of end wall reflection

Peng Dai[*a], Zheng Chen [b], Xiaohua Gan [a], Mikhail A. Liberman [c]

[a] Department of Mechanics and Aerospace Engineering, Southern University of Science and Technology, Shenzhen 518055, China

[b] SKLTCS, CAPT, BIC-ESAT, College of Engineering, Peking University, Beijing 100871, China

[c] Nordita, KTH Royal Institute of Technology and Stockholm University, Roslagstullsbacken 23, SE-10691 Stockholm, Sweden

**Abstract**

The advancement of highly boosted internal combustion engines (ICEs) with high thermal efficiency is mainly constrained by knock and super-knock respectively due to the end gas autoignition and detonation development. At the end of the compression stroke, the pressure wave propagation and reflection in a small confined space may greatly influence the end gas autoignition, leading to different autoignition characteristics from those in a large or open space. The present study investigates the transient autoignition process in an iso-octane/air mixture inside a closed chamber under engine-relevant conditions. The emphasis is given to the assessing effects of the pressure wave-wall reflection and the mechanism of extremely strong pressure oscillations typical for super-knock. It is found that the hot spot induced autoignition in a closed chamber can be greatly affected by shock/pressure wave reflection from the end wall. Different autoignition modes respectively from the hot spot and end wall reflection are identified. A non-dimensional parameter quantifying the interplay between different length and time scales is introduced, which helps to identify different autoignition regimes including detonation development near the end wall. It is shown that detonation development from the hot spot may cause super-knock with devastating pressure oscillation. However, the detonation development from the end wall can hardly produce pressure oscillation strong enough for the super-knock. The obtained results provide a fundamental insight into the knocking mechanism in engines under highly boosted conditions.

## 1. Introduction

Downsized and turbocharged spark ignition engines (SIEs) have attracted increasing interest due to their advantages of higher thermal efficiency and lower fuel consumption. However, the tendency of knock and super-knock is greatly enhanced in highly boosted environment [1, 2], and this remains a major limitation for the development of modern internal combustion engines (ICEs). While conventional knock has direct relevance to the end gas autoignition, super-knock is found to be closely associated with the detonation development induced by the randomly localized hot spot [3-5]. However, the detailed mechanism of super-knock with destructive in-cylinder pressure oscillations is still not well understood. Usually knock and super-knock occur inside a small confined space at the end of the compression stroke. In a closed and relatively small chamber, propagation and reflection of pressure waves form the chamber walls may greatly influence the end gas autoignition leading to autoignition characteristics different from those in a large chamber or in an open space. Therefore, understanding of autoignition and detonation development in a confined space under engine-relevant conditions is of fundamental interest.

A number of studies have been conducted to understand autoignition induced by thermal and/or composition non-uniformity in reactive mixtures. Zel'dovich and co-workers [6, 7] first analyzed different modes of autoignition wave propagation caused by the reactivity gradient. Bradley and co-workers [8, 9] further investigated autoignition induced by a hot spot and identified a detonation peninsula in the plot of two non-dimensional parameters, namely the normalized temperature gradient ($\xi$) and the ratio of acoustic time to excitation time ($\varepsilon$). This detonation peninsular was widely utilized in the studies on knock and super-knock [3, 10-13]. Besides, Liberman and co-workers [14, 15] studied the combustion regimes induced by the initial temperature gradient. They found that the temperature gradient initiated a detonation predicted by detailed chemical models greatly differs from that predicted using a one-step chemistry. Sow et al. [16] analyzed the effects of thermal stratification in bulk end gas on detonation development. In our recent publications [17-22], autoignition and detonation development of large hydrocarbon fuels with negative temperature coefficient (NTC) were investigated. It was found that the low-temperature chemistry greatly complicates the interaction between chemical reaction and pressure wave and that both temperature gradient (i.e. hot spot or cold spot) and concentration gradient can induce the development of detonation under proper conditions.

Previous studies have been focused mainly on the effects of local reactivity non-uniformity and bulk mixture properties on end gas autoignition and detonation development. However, only a few studies considered the influence of pressure wave propagation and reflection inside a small confined space. Yu et al. [23, 24], Pan et al. [25, 26] and Terashima et al. [27, 28] investigated the interaction between normal flame propagation, end gas autoignition and pressure wave in a closed chamber. However, the effects of wall reflection on autoignition and detonation development from local reactivity non-uniformity have not been systematically investigated. The objectives of this study are three-fold: (1) to investigate the transient autoignition process induced by a hot spot inside a small closed chamber, (2) to quantitatively describe the corresponding autoignition modes by using different non-dimensional parameters, and (3) to assess and interpret the effects of wall reflection and investigate the mechanism of extremely strong pressure oscillation typical for super-knock.

## 2. Numerical model

As the main component of the primary reference fuel (PRF) for gasoline, iso-octane is considered in this study. The reduced PRF mechanism [29] is used in simulations. Its performance in terms of predicting autoignition and flame propagation is demonstrated in the Supplemental Material and Ref. [29].

We consider the one-dimensional transient autoignition process in the stoichiometric iso-octane/air mixture initiated by a hot spot in adiabatic, closed chamber. In the present studies turbulence is not considered. The initial conditions are shown in Fig. 1. The initial hot spot is characterized by the linear temperature distribution:

$$T(t=0,x)=\begin{cases} T_{i,0}+(x-x_0)(dT/dx)_i & \text{for } 0\le x\le x_0 \\ T_{i,0} & \text{for } x_0<x\le L \end{cases} \tag{1}$$

where $x$ is spatial coordinate; $x_0$ is the hot spot size and $L$ is the chamber length; $(dT/dx)_i$ is the temperature gradient within the hot spot, and $T_{i,0}$=900 K is the initial temperature outside the hot spot. It is noted that turbulence might greatly affect the autoignition process when the characteristic time scales of different physical-chemical processes are comparable, but this is beyond the scope of the current work.

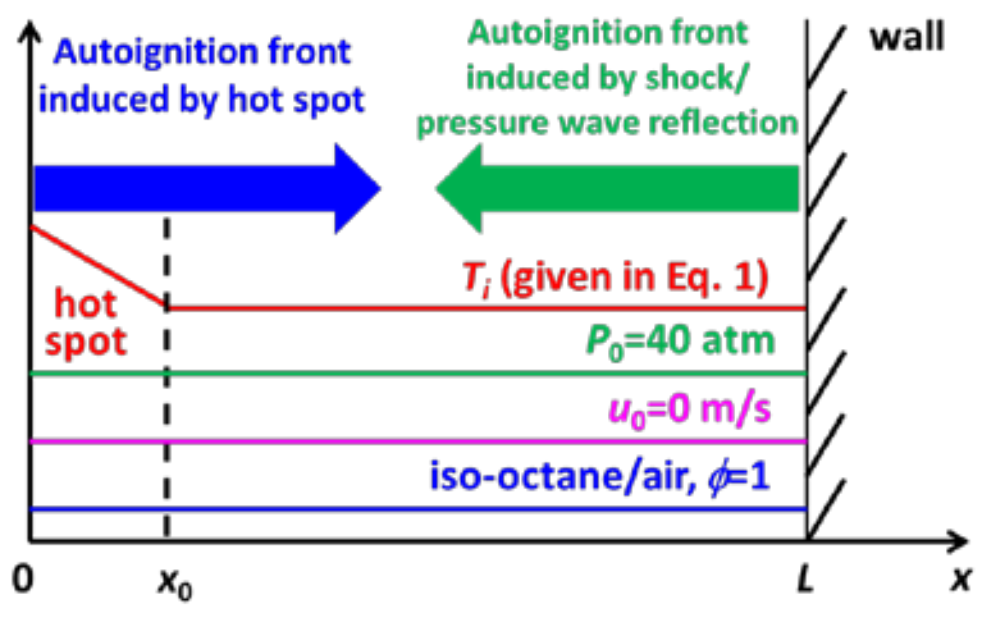

Fig. 1 Schematic of autoignition initiated by a hot spot in iso-octane/air mixture inside a closed planar chamber.

The transient autoignition process is simulated using the in-house code A-SURF [18, 30] which solves the conservation equations for 1D, adiabatic, multi-component, reactive flow using the finite volume method. A multi-level, dynamically adaptive mesh refinement algorithm is adopted to ensure adequate numerical resolution of the reaction zone propagation, pressure/shock wave, and detonation, which are always covered by the finest mesh of width 3.125 μm. The time step is $3.125\times10^{-10}$ s. Details on the governing equations, numerical scheme, and grid convergence can be found in Refs. [17, 18, 21, 30] and thereby are not repeated here.

## 3. Results and discussion

### 3.1 Typical autoignition cases

As shown in Fig. 1, there can be two types of autoignition process happening in a small chamber: (1) a propagating to the right autoignition front induced by the hot spot, and (2) propagating to the left from the end wall the autoignition front induced by the shock/pressure wave reflection.

The autoignition front propagation initiated by a hot spot in large or open space has been extensively studied [8, 9, 14, 17-22]. According to the Zel'dovich theory [7] the critical condition for detonation development by the temperature gradient is defined as the condition of coupling of the reaction wave and the pressure wave. It was shown by Liberman and co-workers [31] that the temperature gradient must be sufficiently shallow such that when the spontaneous wave velocity initiated by the temperature gradient decreases while it propagates along the gradient and reaches its minimum value (the point close to the crossover temperature), it is caught up with the pressure wave which was generated behind the high-speed spontaneous wave front. This condition can be presented by Eq.(2), which is equality of the local sound speed and spontaneous wave speed. We will use it in a slightly different form it is defined as [8, 9]

$$\left(dT/dx\right)_c = \left(a\left(d\tau/dT_0\right)\right)^{-1}, \tag{2}$$

where $\tau$ is the ignition delay time (defined as the time for maximum heat release rate). There is a critical temperature gradient at which the autoignition front propagation speed, $u_a$, is equal to the sound speed, $a$; and it is defined as [8, 9]. The values of ignition delay time $\tau$, excitation time ($\tau_e$, defined as the time interval between 5% and maximum heat release rate [8]) and $(dT/dx)_c$ are shown in Fig. S2 in the Supplemental Material. The normalized temperature gradient of the hot spot, $\xi$, is defined as [8, 9]:

$$\xi = \left(dT/dx\right)_i \Big/ \left(dT/dx\right)_{c,x_0/2} \tag{3}$$

where the subscript $x_0/2$ denotes that the value of critical temperature gradient is evaluated at $x=x_0/2$ in order to represent the average condition within the hot spot. The speed of the autoignition front propagation, $u_a$, can thereby be expressed as [8, 9]:

$$u_a = a/\xi \tag{4}$$

The transient autoignition process at different hot spot sizes ($x_0$), temperature gradients ($\xi$) and chamber lengths ($L$) is systematically investigated. Nine typical cases discussed below with the corresponding details summarized in Table S1 in the Supplemental Material.

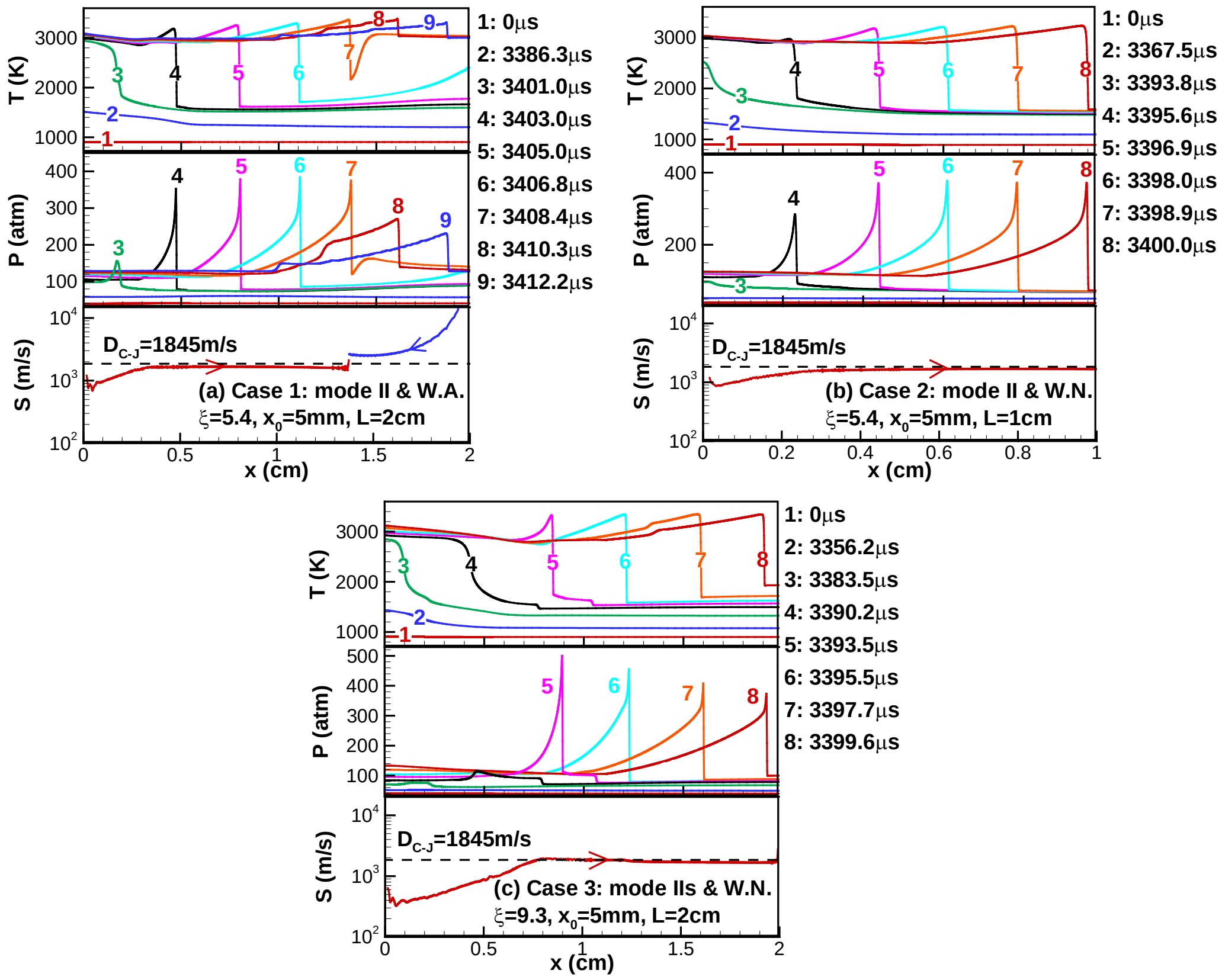

Fig. 2 Temporal evolution of temperature and pressure profiles and transient autoignition front propagation speed, *S*, for (a) case 1, (b) case 2, and (c) case 3. The arrows in the bottom sub-figures denote the propagation direction of autoignition fronts. The horizontal dashed lines indicate the C-J detonation wave speed, $D_{C\text{-}J}$.

Figure 2 shows the autoignition processes for three typical cases: case 1 ($\xi$=5.4, $x_0$=5 mm, $L$=2 cm) for reference, case 2 ($\xi$=5.4, $x_0$=5 mm and $L$=1 cm) with shorter chamber length, and case 3 ($\xi$=9.3, $x_0$=5 mm and $L$=2 cm) with larger $\xi$. For autoignition initiated by the hot spot, three modes occur sequentially with the increase of $\xi$, namely: (I) supersonic reaction front propagation, (II) detonation development, and (III) subsonic reaction front propagation. For cases 1 and 2 shown in Figs. 2(a) and 2(b), a detonation wave develops inside the hot spot and thereby they correspond to autoignition mode II. The detonation development is caused by strong coherent coupling and mutual amplification between chemical reaction and pressure wave. On the other hand, Fig. 2(c) shows that for higher value of $\xi$, a leading shock wave is generated ahead of the autoignition front, which is the result of weaker chemical-acoustic interaction at reduced autoignition front propagation speed (see Eq. 4) [17, 18]. The autoignition front behind the shock wave gradually develops to a detonation wave due to the compression and heating of upstream mixture by the shock wave. This autoignition mode was also observed in Ref. [17, 18] and it is referred as mode IIs ("s" denotes "shock wave") in this study.

On the other hand, line #6 in Fig. 2(a) indicates that autoignition occurs near the end wall before arrival of the detonation wave and that it produces a supersonic autoignition front propagating to the left. This autoignition mode near the end wall is thereby referred as W.A. (Wall Autoignition). The propagating to the right detonation wave quenches after encountering the propagating to the left autoignition front since all the reactants are consumed; and it degenerates to a shock wave propagating to the right (see lines #7-9 in Fig. 2a).

When the chamber length is reduced as in case 2, no autoignition near the end wall is observed. This is mainly because the time for the detonation wave propagation is shorter than the autoignition time near the end wall. This mode near the wall is hence referred as W.N. (Wall without autoignition). The mode W.N. is also observed in case 3 with larger $\xi$. Therefore, for both cases 2 and 3, the detonation wave can reach the end wall and their shock wave reflection is much stronger than that in case 1. This is demonstrated in Fig. 3, which shows the temporal evolution of pressure at the end wall for the cases 1, 2 and 3.

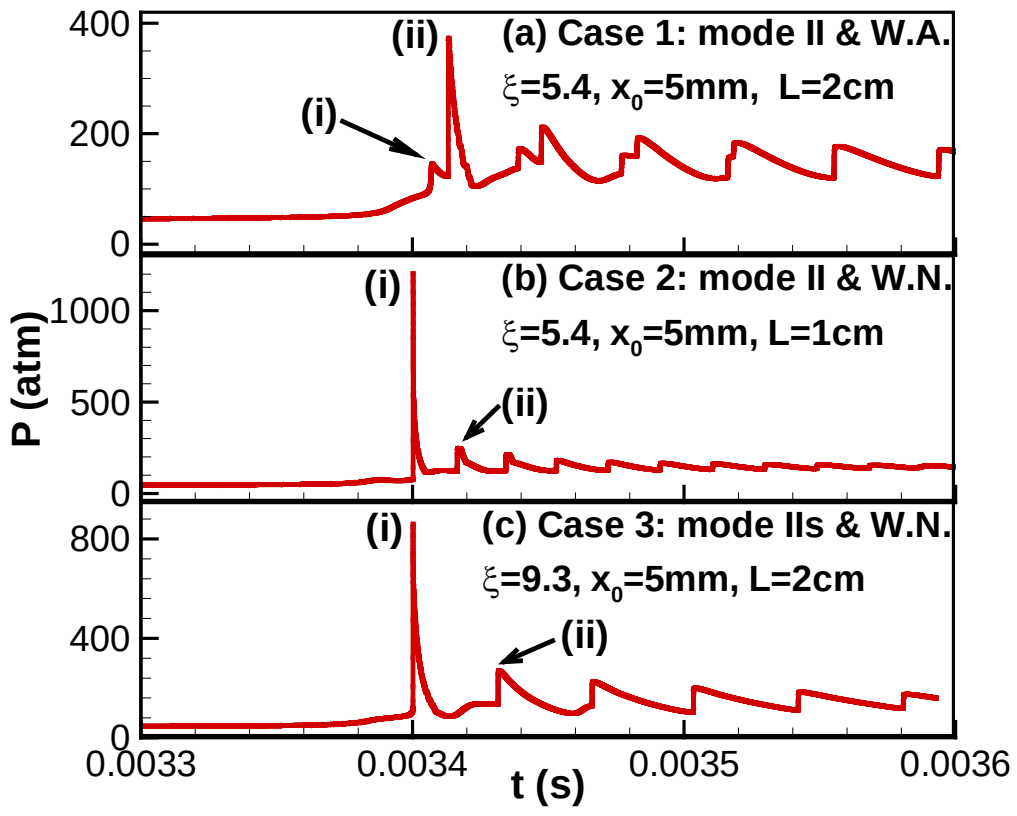

Fig. 3 Temporal evolution of pressure at the end wall for (a) case 1, (b) case 2, and (c) case 3. The marked pressure peaks correspond to: (i) autoignition near the end wall, (ii) reflection of the shock wave remaining from the quenching detonation wave in Sub-figure (a); and (i) reflection of detonation wave, (ii) reflection of shock wave after quenching the reflected detonation wave in Sub-figures (b) and (c).

The pressure oscillation in Fig. 3 is caused by the back-and-forth propagation of detonation/shock/pressure waves in the closed chamber. It is seen that the maximum pressure and hence the maximum pressure oscillation amplitude in cases 2 and 3 are much higher than those in the case 1. As shown in Fig. 2, this is because the detonation waves in cases 2 and 3 directly hit against the end wall and generate extremely high pressure via reflection, while the detonation wave in case 1 degenerates to a shock wave before reaching the end wall. Therefore, a combination of mode II or IIs from the hot spot and mode W.N. near the end wall results in the strongest pressure oscillation and most severe damage to the engine cylinder. The extreme high pressure oscillation in cases 1, 2 and 3 are respectively around 270 atm, 1130 atm, and 780 atm, which may all correspond to the super-knock phenomenon in SIEs.

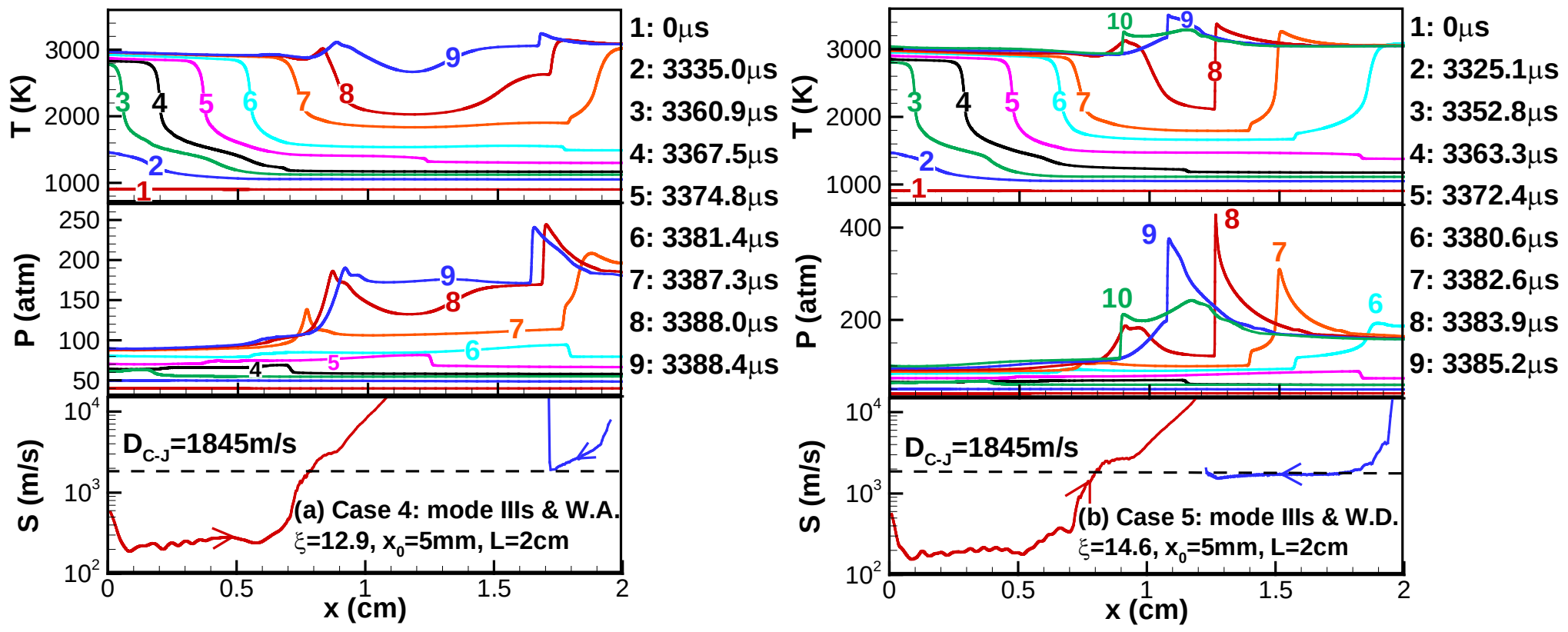

Fig. 4 Temporal evolution of temperature and pressure profiles and transient autoignition front propagation speed for (a) case 4 and (b) case 5.

Figure 4 shows the autoignition processes in case 4 ($\xi$=12.9, $x_0$=5 mm and $L$=2 cm) and case 5 ($\xi$=14.6, $x_0$=5 mm and $L$=2 cm) with further increase in the value of $\xi$. It is observed that the autoignition mode from the hot spot transits to subsonic reaction front propagation with a leading shock wave, which is referred as mode IIIs ("s" denotes "shock wave"). On the other hand, the mixture near the end wall autoignites in both cases 4 and 5 due to increased pressure in the reflected shock wave. However, while the supersonic autoignition front propagating to the left is generated in case 4 (i.e. mode W.A.), a detonation wave is formed from the end wall in case 5, which is referred as mode W.D. (Wall Detonation). The detonation wave quenches when it encounters the autoignition wave propagating to the right degenerates to a shock wave that propagates to the left side in the burnt mixture (see lines #9 and #10 in Fig. 4b).

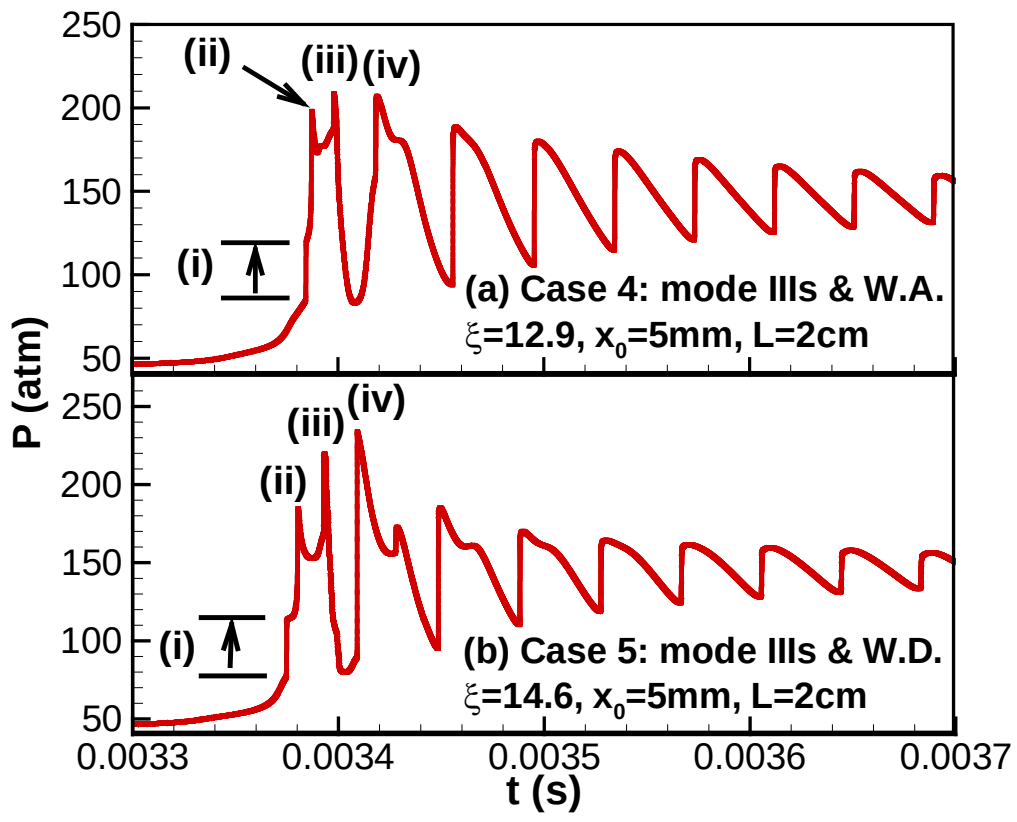

Fig. 5 Temporal evolution of pressure at the end wall for (a) case 4 and (b) case 5. The marked pressure peaks correspond to: (i) reflection of leading shock wave, (ii) autoignition near the end wall, (iii) reflection of pressure wave formed by the autoignition front from the hot spot, and (iv) reflection of shock wave formed by (a) the supersonic autoignition front and (b) the quenching detonation wave from the end wall.

Figure 5 shows the temporal evolution of pressure at the end wall in cases 4 and 5. It is observed that although a detonation wave is formed from the end wall in case 5, the corresponding pressure oscillation amplitude is very close to that in case 4 without detonation development, which are both much lower than those in cases 1-3 with detonation wave initiated from the hot spot. This is because the shock wave propagating to the left formed by the quenching detonation wave in case 5 has to propagate a round trip (i.e. back and forth through the chamber) before arriving at the end wall, which significantly weakens its strength. Therefore, the detonation wave generated from the end wall (i.e. mode W.D.) may not cause damage as severe as the detonation wave generated by the hot spot (i.e. modes II and IIs)

The autoignition process in case 6 ($\xi$=16.2, $x_0$=5 mm and $L$=2 cm) with further increased $\xi$ is shown in the Supplemental Material. It is noted that the corresponding autoignition mode is essentially the same as that in cases 4 (i.e. mode IIIs & W.A.) and a detonation wave is not developed. Besides cases 1-6 with $x_0$=5 mm, we also consider cases with smaller hot spot of $x_0$=3.5 mm, including case 7 ($\xi$=5.4, $L$=2 cm, mode II & W.A.), case 8 ($\xi$=9.5, $L$=2 cm, mode IIs & .W.A.), and case 9 ($\xi$=14.8, $L$=2 cm, mode IIIs & W.A.). The corresponding autoignition processes are presented in the Supplemental Material.

**3.2 Map of various autoignition modes**

The above results indicate that the autoignition in a small chamber is essentially a combination of the autoignition initiated by the hot spot and that induced by shock/pressure wave reflection on the end wall.

For the autoignition from the hot spot, five modes can be sequentially identified with the increase of $\xi$, namely: supersonic reaction front propagation (I), detonation development without leading shock wave (II), detonation development with leading shock wave (IIs), subsonic reaction front propagation with leading shock/pressure wave (IIIs), and subsonic reaction front propagation without leading shock/pressure wave (III). Figure 6(a) shows the corresponding regimes of these autoignition modes in $\xi$-$\varepsilon$ diagram. The non-dimensional parameter, $\varepsilon$, is defined as the ratio of the acoustic time, $x_0/a$, and the excitation time, $\tau_e$ (i.e. $\varepsilon=x_0/(a\tau_e)$) [8]. It is noted that the strength of the leading shock/pressure wave in regime IIIs decreases with $\xi$. That is because the speed of the subsonic reaction front decreases with increasing $\xi$ and thereby the coherent chemical-acoustic interaction is weakened. When $\xi$ is sufficiently large, no shock/pressure wave is generated within the hot spot and the autoignition mode transits from mode IIIs to mode III.

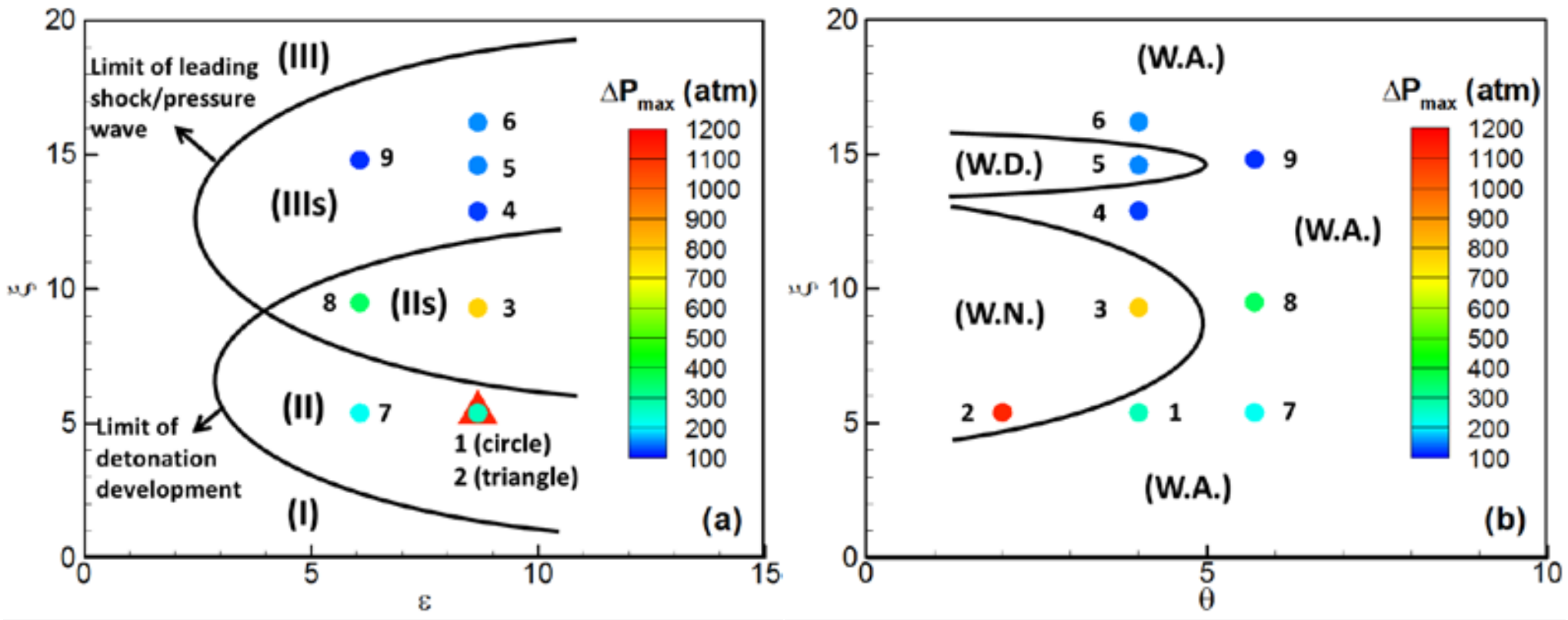

Fig. 6 Regimes of autoignition modes which are respectively (a) initiated by the hot spot and (b) induced by shock/pressure wave reflection on the end wall. Autoignition cases 1 to 9 are also plotted with the corresponding maximum pressure oscillation amplitude at the end wall, $\Delta P_{max}$, indicated by the color scale.

On the other hand, three autoignition modes can be identified near the end wall, which are respectively: autoignition with supersonic reaction front propagation (W.A.), non-autoignition (W.N.), and autoignition with detonation development (W.D.). There are four important factors affecting the autoignition mode near the end wall:

(1) The time elapsed between the autoignition in the hot spot and that at the end wall, which can be represented by the difference between 0D ignition delay at the hot spot ($x$=0) and that at the end wall ($x$=$L$):

$$\Delta\tau_{0D} = \tau_{0D,\,wall} - \tau_{0D,\,hot\ spot} \tag{5}$$

According to Eqs. (1)-(3), we have

$$\begin{aligned}\Delta\tau_{0D} &= (d\tau/dT_0)(T_{i,wall} - T_{i,hot\ spot}) = (d\tau/dT_0)(dT/dx)_i\, x_0 \\ &= (d\tau/dT_0)\left(\xi (dT/dx)_c\right) x_0 = \xi x_0/a\end{aligned} \tag{6}$$

(2) The autoignition front from the hot spot propagates to the end wall during time:

$$t_a = L/S \sim L/u_a = L\xi/a \tag{7}$$

(3) The leading shock/pressure wave from the hot spot propagates to the end wall during

time:

$$t_{wave} \sim L/a \tag{8}$$

(4) The strength of the leading shock/pressure wave, $I_{wave}$.

It is convenient to introduce a non-dimensional parameter,

$$\theta = L/x_0 \tag{9}$$

to characterize the interplay between the length scales $L$ and $x_0$ as well as that between the time scales $t_a$ and $\Delta\tau_{0D}$. Substituting Eqs. (6) and (7) into (9) yields

$$\theta = (L\xi/a)/(\xi x_0/a) \sim t_a/\Delta\tau_{0D} \tag{10}$$

Figure 6(b) shows the autoignition regimes near the end wall in $\xi$-$\theta$ diagram. The transition of autoignition modes with increasing $\xi$ for cases 1, 3-6 will be explained as below.

(1) The transition from the mode W.A. to modes W.N. and W.N. to W.A. for $\xi$<13 is a result of competition between increasing $\Delta\tau_{0D}$ and $t_a$. When $\xi$ is relatively small, the increase of $\xi$ leads to a longer $\Delta\tau_{0D}$ (see Eq. 6) which dominates and delays the autoignition near the end wall relative to that at the hot spot. Therefore, the autoignition transits from mode W.A. in case 1 to mode W.N. in case 3. However, with further increase of $\xi$, the increase in $t_a$ (see Eq. 7) dominates and provides more time for the mixture near the end wall to autoignite before the arrival of the autoignition front from the hot spot. This leads to the further transition from mode W.N. in case 3 to mode W.A. in case 4.

(2) The transition from the mode W.A. to W.D. and W.D. to W.A. for $\xi$>13 is a result of competition between increasing $t_a$ and decreasing $I_{wave}$. At the beginning, an increase in $\xi$ leads to further increase in $t_a$, so that the autoignition front initiated from the end wall has more time to form a detonation wave behind the reflected shock/pressure wave. Therefore, the autoignition mode transits from W.A. in case 4 to W.D. in case 5. However, when $\xi$ further increases, the strength of leading shock/pressure wave, $I_{wave}$, decreases to the extent that the reflected shock/pressure wave from the wall can no longer initiate a detonation wave. Therefore, the autoignition mode transits from W.D. in case 5 to W.A. in case 6.

According to Eq. (10), lower vaule of $\theta$ implies less time needed for the autoignition front developed from the hot spot to propagate to the end wall (i.e. shorter $t_a$) or longer time for the mixture near the end wall to autoignite after the hot spot autoignition (i.e. longer $\Delta\tau_{0D}$), either of which reduces the tendency of autoignition near the end wall and is favorable to mode W.N. Therefore, the regime of mode W.N. expands with decreasing $\theta$ and lies within a reversed C-shaped curve in $\xi$-$\theta$ diagram.

As for mode W.D., according to Eqs. (6) and (8), the ratio between the time scales $\Delta\tau_{0D}$ and $t_{wave}$ is:

$$\Delta\tau_{0D}/t_{wave} \sim (\xi x_0/a)/(L/a) = \xi/\theta \qquad (11)$$

Higher value of ($\Delta\tau_{0D}/t_{wave}$) indicates that either the autoignition front initiated from the end wall has more time to form detonation wave before thermal explosion throughout the chamber (i.e. longer $\Delta\tau_{0D}$) or the mixture near the end wall is earlier compressed by the reflected shock/pressure wave (i.e. shorter $t_{wave}$). Therefore, increase of ($\xi/\theta$) promotes the tendency of detonation development from the end wall, which implies that the slop of the lower limit of W.D. regime is positive in $\xi$-$\theta$ diagram.

On the other hand, either smaller $x_0$ (i.e., larger $\theta$, see Eq. 9) or larger $\xi$ around W.D. regime weakens the chemical-acoustic interaction within the hot spot and thereby leads to lower strength of leading shock/pressure wave, $I_{wave}$, which reduces the tendency of detonation development from the end wall. Therefore, the slop of the upper limit of W.D. regime is negative. Summarizing, W.D. regime lies within a reversed C-shaped curve in $\xi$-$\theta$ diagram.

Figure 6 provides fundamental insights into the condition for both hot spot induced autoignition modes in a small chamber and super-knock in ICEs, which can be quantitatively described in $\xi$-$\varepsilon$ and $\xi$-$\theta$ diagrams. It is noted that the strongest pressure oscillation occurs in modes II & W.N. and IIs & W.N. (e.g. cases 2 and 3 with the maximum pressure oscillation amplitude above 750 atm), which correspond to the detonation wave from the hot spot directly hitting the end wall. Besides, modes II and IIs with detonation development from the hot spot (e.g. cases 1, 2, 3, 7 and 8 with the maximum pressure oscillation amplitude above 200 atm) may cause super-knock while mode W.D. with detonation development from the end wall (e.g. case 5) can hardly produce pressure oscillation strong enough for super-knock. It is also emphasized that the correspondence between the autoignition modes from the hot spot (see Fig. 6a) and those near the end wall (see Fig. 6b) is not limited by the cases discussed in this study. The autoignition regimes in $\xi$-$\varepsilon$ and $\xi$-$\theta$ diagrams may quantitatively depend on different factors including the initial temperature and pressure, mixture composition, hot spot (size, temperature gradient), chamber size, and chamber geometric (planar, cylindrical or spherical), which deserves further investigation.

**4. Conclusions**

The 1D transient autoignition process initiated by a hot spot in a stoichiometric iso-octane/air mixture within a small closed chamber is numerically investigated considering detailed chemistry and transport. The effects of shock/pressure wave propagation and reflection on the end wall are systematically investigated. It is found that the autoignition in a small chamber consists of two parts, i.e., (1) the autoignition front induced by the hot spot and propagating to the right, and (2) the autoignition front from the end wall induced by the temperature rise due to shock/pressure wave reflection there and propagating to the left. Five autoignition modes from the hot spot are identified and quantitatively described in $\xi$-$\varepsilon$ diagram. Three autoignition modes from the end wall including detonation development are also identified. A new non-dimensional parameter, $\theta$, is introduced to quantify the interplay between different length and time scales. The autoignition modes from the end wall are quantified in $\xi$-$\theta$ diagram. A combination of $\xi$-$\varepsilon$ and $\xi$-$\theta$ diagrams gives a quantitative prediction of autoignition modes caused by a hot spot in a small closed chamber. Besides, it is demonstrated that detonation developed from the hot spot may cause super-knock while detonation developed from the end wall can hardly produce pressure oscillation strong enough for super-knock. Mechanisms of different autoignition modes can be interpreted with the help of $\xi$-$\varepsilon$ and $\xi$-$\theta$ diagrams. Nevertheless, further investigation on the influence of initial thermal conditions on the autoignition regimes in $\xi$-$\varepsilon$ and $\xi$-$\theta$ diagrams is still needed.

**Acknowledgments**

This work is supported by National Natural Science Foundation of China (Nos. 51976088 and 51606091).

**List of Figure captions**

**Fig. 1** Schematic of autoignition initiated by a hot spot in iso-octane/air mixture inside a closed planar chamber.

**Fig. 2** Temporal evolution of temperature and pressure profiles and transient autoignition front propagation speed, *S*, for (a) case 1, (b) case 2, and (c) case 3. The arrows in the bottom sub-figures denote the propagation direction of autoignition fronts. The horizontal dashed lines indicate the C-J detonation wave speed, $D_{C\text{-}J}$.

**Fig. 3** Temporal evolution of pressure at the end wall for (a) case 1, (b) case 2, and (c) case 3. The marked pressure peaks correspond to: (i) autoignition near the end wall, (ii) reflection of shock wave formed by the quenching detonation wave in Sub-figure (a); and (i) reflection of detonation wave, (ii) reflection of shock wave formed by the quenching reflected detonation wave in Sub-figures (b) and (c).

**Fig. 4** Temporal evolution of temperature and pressure profiles and transient autoignition front propagation speed for (a) case 4 and (b) case 5.

**Fig. 5** Temporal evolution of pressure at the end wall for (a) case 4 and (b) case 5. The marked pressure peaks correspond to: (i) reflection of leading shock wave, (ii) autoignition near the end wall, (iii) reflection of pressure wave formed by the autoignition front from the hot spot, and (iv) reflection of shock wave formed by (a) the supersonic autoignition front and (b) the quenching detonation wave from the end wall.

**Fig. 6** Regimes of autoignition modes which are respectively (a) initiated by the hot spot and (b) induced by shock/pressure wave reflection on the end wall. Autoignition cases 1 to 9 are also plotted with the corresponding maximum pressure oscillation amplitude at the end wall, $\Delta P_{max}$, indicated by the color scale.